\documentclass[conference,letterpaper]{IEEEtran}

\IEEEoverridecommandlockouts

\usepackage[utf8]{inputenc}
\usepackage[T1]{fontenc}
\usepackage{cite}
\usepackage{graphicx}
\usepackage{psfrag}
\usepackage{subfigure}
\usepackage[caption=false]{subfig}
\usepackage{url}
\usepackage{ifthen}
\usepackage[cmex10]{amsmath}
\usepackage{array}
\usepackage{amssymb}
\usepackage{amsthm}
\usepackage{mathtools}
\usepackage{amsfonts}
\usepackage{graphicx}
\usepackage{epstopdf}
\usepackage{algorithm}
\usepackage{algorithmic}

\newtheorem{lemma}{Lemma}

\newtheorem{theorem}{Theorem}

\usepackage{setspace}
\usepackage{amscd}
\usepackage{mathrsfs}
\usepackage{epsfig}
\usepackage{color}
\usepackage{textcomp}
\usepackage{comment}
\usepackage{xcolor}
\usepackage{pgf}
\usepackage{pgfplots}
\pgfplotsset{compat=1.9}
\usepackage{tikz}
\usetikzlibrary{decorations.pathreplacing}
\usetikzlibrary{decorations.markings}
\usetikzlibrary{patterns}
\usepackage[outline]{contour}
\contourlength{1.2pt}
\usetikzlibrary{plotmarks}
\usetikzlibrary{spy}
\usetikzlibrary{calc}
\usetikzlibrary{shapes,plotmarks,positioning,fit,chains,tikzmark,arrows,hobby,arrows.meta}
\usetikzlibrary{decorations.pathreplacing,calligraphy}
\usepgfplotslibrary{groupplots}

\usepackage{import}

\IEEEoverridecommandlockouts

\usepackage[toc, notranslate, acronym]{glossaries}
\usepackage{glossaries-extra}
\setabbreviationstyle[acronym]{long-short}
\newacronym{4g}{4G}{fourth generation}
\newacronym{5g}{5G}{fifth generation}
\newacronym{6g}{6G}{sixth generation}
\newacronym{iid}{i.i.d.}{independent and identically distributed}
\newacronym{llm}{LLM}{large language model}
\newacronym{kl}{KL}{Kullback Leibler}
\newacronym{occ}{OCC}{online conformal compression}
\newacronym{cp}{CP}{conformal prediction}
\newacronym{ocp}{OCP}{online conformal prediction}
\newacronym{ai}{AI}{artificial intelligence}
\newacronym{bcc}{BCC}{block conformal compression}

\DeclareMathOperator*{\argmax}{argmax}

\newcommand{\Cbrac}[1]{\left\{#1\right\}}

\newcommand{\EEE}{\mathrm{e}}

\newcommand{\R}{\mathbf{R}}

\newcommand{\XXX}{\mathcal{X}}

\newcommand{\one}{\mathbbm{1}}

\usepackage{dsfont}
\usepackage{bbm}

\usepackage{siunitx-v2}
\begin{document}

\title{Online Conformal Compression for Zero-Delay Communication with Distortion Guarantees
    \thanks{The work of Unnikrishnan Kunnath Ganesan and Giuseppe Durisi was supported in part by the Swedish Foundation for Strategic Research,
    under the project SAICOM. 
    The work of Matteo Zecchin and Osvaldo Simeone was supported by the European Union’s Horizon Europe project CENTRIC (101096379). The work of Osvaldo Simeone was also supported by the Open Fellowships of the EPSRC (EP/W024101/1) and by the EPSRC project  (EP/X011852/1).
    The work of Petar Popovski was supported, in part, by the Villum Investigator Grant “WATER” from the Velux Foundation, Denmark.
    }
}

\author{%
	\IEEEauthorblockN{Unnikrishnan Kunnath Ganesan\IEEEauthorrefmark{1},
                          Giuseppe Durisi\IEEEauthorrefmark{1}, 
                          Matteo Zecchin\IEEEauthorrefmark{2}, 
                          Petar Popovski\IEEEauthorrefmark{3}, and 
                          Osvaldo Simeone\IEEEauthorrefmark{2}  }
	\IEEEauthorblockA{\IEEEauthorrefmark{1}
            Department of Electrical Engineering,
		Chalmers University of Technology,
		412 96 Göteborg, Sweden\\
		Email: \{kunnathg, durisi\}@chalmers.se}
	\IEEEauthorblockA{\IEEEauthorrefmark{2}
            CIIPS, Department of Electrical Engineering,
		Kings College London, Strand, London, WC2R 2LS United Kingdom \\
		Email: \{matteo.1.zecchin,osvaldo.simeone\}@kcl.ac.uk}
	\IEEEauthorblockA{\IEEEauthorrefmark{3}
            Department of Electronic Systems,
		Aalborg University,
		9220 Aalborg, Denmark\\
		Email: petarp@es.aau.dk}
}




\maketitle

\begin{abstract}
We investigate a lossy source compression problem in which both the encoder and decoder are equipped with a pre-trained sequence predictor.
We propose an online lossy compression scheme that, under a $0-1$ loss distortion function, ensures a deterministic, per-sequence upper bound on the distortion (outage) level for any time instant.
The outage guarantees apply irrespective of any assumption on the distribution of the sequences to be encoded or on the quality of the predictor at the encoder and decoder.
The proposed method, referred to as \gls{occ}, is built upon online conformal prediction---a novel method for constructing confidence intervals for arbitrary predictors.
Numerical results show that \gls{occ} achieves a compression rate comparable to that of an idealized scheme in which the encoder, with hindsight, selects the optimal subset of symbols to describe to the decoder, while satisfying the overall outage constraint.
\end{abstract}

\glsresetall


\section{Introduction}
\subsection{Context and Motivation}
Advances in artificial intelligence have made autoregressive sequence predictors, based on models such as transformers or state-space models, widely available.
It is well known that sequence predictors can be directly leveraged for \emph{lossless} compression via standard mechanisms such as arithmetic coding~\cite{valmeekam2023llmzip,deletang2024language}.
The goal of this work is to introduce for the first time a zero-delay compression methodology that guarantees a \emph{deterministic}, per sequence, upper bound on the distortion, when the latter is measured via a $0-1$ loss.

\begin{figure*}
	\centering
	\newdimen\R
	\R=1cm
	\begin{tikzpicture}[auto]

    \node[draw, rectangle, minimum width=\R, minimum height=\R, align=center] (est_symb1) {$\hat{X}_1, \hat{X}_2,\dots, \hat{X}_{t-1}$};

    \node[draw, rectangle, minimum width=\R, minimum height=\R, align=center, right=of est_symb1] (predictor1) {$\mu_t(X_t|\hat{X}^{t-1})$\\predictor};

    \node[draw, rectangle, minimum width=\R, minimum height=\R, align=center, below=of predictor1] (cp1) {conformal prediction};

    \node[draw, rectangle, minimum width=2*\R, minimum height=\R, align=center, below=of cp1] (encoder) {encoder};

    \node[draw, rectangle, minimum width=2*\R, minimum height=\R, align=center, right=of encoder] (channel) {noiseless\\channel};

    \node[draw, rectangle, minimum width=2*\R, minimum height=\R, align=center, right=of channel] (decoder) {decoder};

    \node[align=center, left=2*\R of encoder] (input) {};
    \draw[->] (input) -- (encoder) node[midway, above] {$X_t \in \XXX$} node[midway, below] {true symbol};

    \node[align=center, below=4.5*\R of est_symb1] (dummy) {};

    \draw[->] (est_symb1) -- (predictor1);
    \draw[->] (predictor1) -- (cp1) ; 
    \draw[->] (cp1) -- (encoder) node[midway, left] {$\hat{\XXX}_t$};

    \draw[->,dashed] (encoder.south) |- (dummy.south) -- (est_symb1.south) node[midway,yshift=1*\R, left] {$\hat{X}_t$};

    \draw[->,dashed] ([yshift=-1.52*\R]est_symb1.south) -- (cp1.west) ;

    \node [draw,dotted,fit=(est_symb1) (predictor1) (cp1)]{};

    \node[align=center, right=\R of decoder] (output) {};

    \draw[->] (encoder) -- (channel) node[midway, above] {$M_t$};
    \draw[->] (channel) -- (decoder);

    \node[draw, rectangle, minimum width=\R, minimum height=\R, align=center, above=of decoder] (cp2) {conformal prediction};
    \draw[->] (cp2) -- (decoder) node[midway, right] {$\hat{\XXX}_t$};

    \node[draw, rectangle, minimum width=\R, minimum height=\R, align=center, above=of cp2] (predictor2) {$\mu_t(X_t|\hat{X}^{t-1})$\\predictor};
    
    \node[draw, rectangle, minimum width=\R, minimum height=\R, align=center,right=of predictor2] (est_symb2) {$\hat{X}_1, \hat{X}_2,\dots, \hat{X}_{t-1}$};
    \draw[->] (est_symb2) -- (predictor2);
    \draw[->] (predictor2) -- (cp2) ; 
    \node [draw,dotted,fit=(est_symb2) (predictor2) (cp2)]{};
    \draw[->] (decoder) -| (est_symb2) node[midway, right] {$\hat{X}_t$};

    \draw[->,dashed] ([yshift=-1.52*\R]est_symb2.south) -- (cp2.east) ;

\end{tikzpicture}
	\caption{Block diagram representing the operation of the proposed online conformal compression (OCC) strategy for zero-delay coding with anytime deterministic outage guarantees.}
	\label{fig:system_model}
\end{figure*}
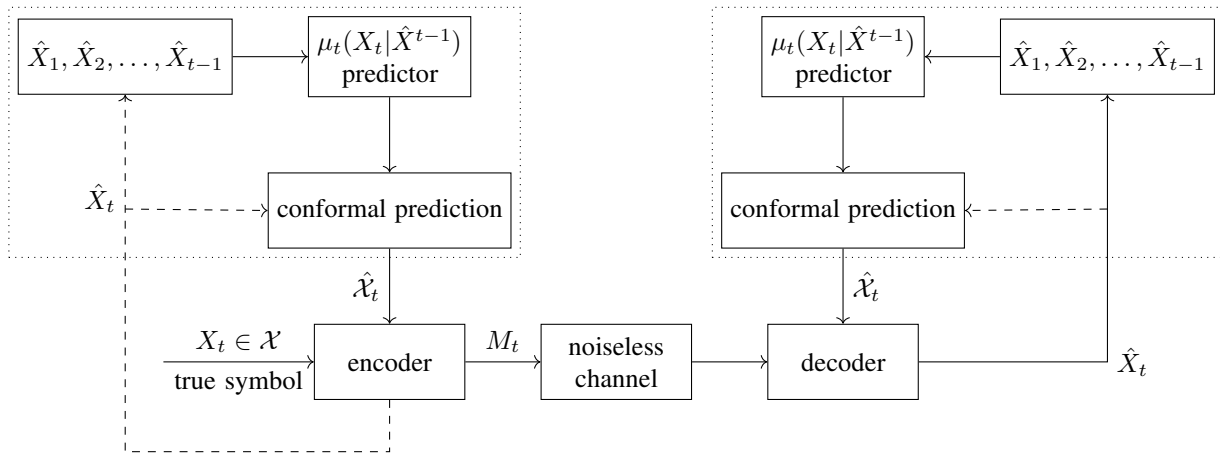

To elaborate, consider the setting in Fig.~\ref{fig:system_model}, in which a sequence
$\{X_t\}$, on whose distribution no assumption is made, is encoded for transmission over a
noiseless channel in a zero-delay, symbol-by-symbol fashion.
Encoder and decoder are equipped with the same autoregressive sequence predictor.
We measure the distortion as the fraction of symbols that are incorrectly recovered at the
receiver, and we wish to ensure a deterministic upper bound on the cumulative distortion
at any time~$t$, while minimizing the coding rate. This requirement is in contrast with
research in the domain of
individual-sequence, universal encoding
methods~\cite{linder2001zero,weissman2002limited}, in which one typically provides
asymptotic guarantees on the expected normalized cumulative distortion,
for a fixed coding rate.
Specifically,~\cite{linder2001zero} proposes a zero-delay
adaptive coding scheme in which a common randomization is utilized at both the encoder and
decoder, ensuring that the maximum
difference between the cumulative distortion and the one achievable by the best scalar
quantizer matched to the sequence, approaches zero as the length
of the encoded sequence increases.
This result was generalized in several directions in~\cite{weissman2002limited}, where the
authors consider arbitrary alphabets and distortion measures, a broader reference class
than just scalar quantizers, and arbitrary finite delay.
They also eliminate the need for common randomness.

This paper proposes a novel zero-delay prediction-aided strategy with anytime, deterministic, guarantees on the accumulated outage rate. The approach is built upon \emph{\gls{ocp}}.
\Gls{cp} has emerged as a robust framework for uncertainty quantification, offering a principled approach to transforming point-based predictions into calibrated prediction sets, thereby combining the strengths of machine learning with the theoretical rigor of statistical methods~\cite{angelopoulos2024theoretical}.
While \gls{cp} relies on the statistical assumption of exchangeability, \gls{ocp} offers
deterministic, worst-case guarantees~\cite{gibbs2024conformal}.
Recent extensions of \gls{ocp} encompass algorithm with Bayesian interpretations~\cite{zhang2024benefit},
and algorithms that have guarantees that are multi-valid and localized in time or
in the covariate space~\cite{bastani2022practical,bhatnagar2023improved,zecchin2024localized}.
Such \gls{ocp} algorithms can be used, for example, to design acquisition
policies in Bayesian optimization~\cite{stanton2023bayesian,kim2024robust} and safe control
policies in robotics~\cite{dixit2023adaptive,lekeufack2024conformal,muthali2023multi}, as
well as
to calibrate machine-learning models in wireless communication
systems~\cite{cohen2023calibrating,cohen2023guaranteed}.

To the best of our knowledge, this is the first work presenting an application of \gls{ocp} to compression.

\subsection{Contributions}
\begin{itemize}
	\item We propose a zero-delay lossy source compression scheme, referred to as \emph{\gls{occ}},  that leverages any autoregressive sequence predictor available at both encoder and decoder.

	\item Under a $0-1$ distortion loss, \gls{occ} is proved to ensure a deterministic anytime upper bound on the outage level that holds irrespective of any assumption on the distribution of the source or on the quality of the predictor.

	\item Numerical results show that \gls{occ} achieves a compression rate comparable to that of an idealized  scheme in which the encoder, with hindsight, selects the optimal subset of symbols to describe to the decoder while satisfying the overall outage constraint.
\end{itemize}

\section{System Model}

As illustrated in Fig.~\ref{fig:system_model}, we consider the problem of prediction-based lossy compression of a sequence $X_1,X_2,\dots,$ taking values in a discrete finite alphabet $\mathcal{X}$. We make no assumption on the joint distribution of the sequence, and
denote by $X^t=[X_1,\dots,X_t]$ the first $t$ symbols of the sequence.

At each time $t$, the encoder applies a zero-delay $t$-dependent encoding function that maps the
currently observed symbol $X_{t}$ to a $b_t$-bit message $M_t$, which is conveyed to
the decoder via a noiseless channel.
As we shall clarify shortly, the message $M_{t}$ may include additional signaling needed
to decode the symbols unambiguously.
Upon receiving the message $M_t$, the decoder applies a zero-delay decoding function to
obtain an estimate $\hat{X}_t$ of the source symbol $X_t$.

The sequence of encoding and decoding functions is designed based on the output of a
sequence predictor. Specifically, we assume that both the encoder and the decoder have
access to a pre-trained sequence predictor which, given a sequence ${X}^{t-1}$,
returns a probability distribution $\mu_t(X_t|{X}^{t-1})$ for the next symbol
$X_t\in\mathcal{X}$.
The predictor is treated as a black box, and is allowed to vary over time, allowing for
possible online fine-tuning. No assumptions are made regarding its predictive performance.

We adopt the $0-1$  distortion function $\one\{\hat{X} \neq X\}$, which returns $1$ if the reconstructed symbol $\hat{X}$ is different from the original symbol $X$ (in which case, an outage event occurs), and returns $0$ otherwise.
Fix a target long-term outage rate $\alpha \in (0,1)$.
The goal  is to design a sequence of encoding and decoding functions to minimize the transmission rate
\begin{align}\label{eqn:avg_compression_rate}
	B_T = \frac{1}{T} \sum_{t=1}^{T} b_t,
\end{align}
while satisfying the worst-case, deterministic,  distortion requirement  \begin{align}
	\label{eqn:misc_guarantee1}
	\left|\frac{1}{T}\sum^T_{t=1}\one{\{\hat{X}_{t}\neq X_{t}\}}-\alpha\right|\leq \frac{C}{T}
\end{align} for some constant $C>0$.
The constraint (\ref{eqn:misc_guarantee1}) amounts to an upper bound on the fraction of symbols that are erroneously decoded.
The bound must apply simultaneously for all times $T$, and it must hold for any sequence $X^T$.
Note that this requirement entails the limit
\begin{align}
	\label{eqn:dist_guarantee}
	\limsup _{T\to\infty} \frac{1}{T} \sum^T_{t=1} \one\{\hat{X}_t \neq X_t\}\leq \alpha.
\end{align}

We address the optimization problem in~\eqref{eqn:avg_compression_rate}--\eqref{eqn:misc_guarantee1} in two different settings, which model \textit{synchronous} and \textit{asynchronous} communication modes, respectively:
\begin{itemize}
	\item \emph{Synchronous transmission}: In the synchronous transmission case, both the
	      encoder and decoder have a shared clock or predefined timing agreement, so that
	      symbols are generated and transmitted at specific time instants known to both
	      parties. This ensures that the decoder is able to  to associate unambiguously the
	      received bits to the corresponding decoded symbols.
	\item \emph{Asynchronous transmission}: In the asynchronous transmission case, encoder
	      and decoder do not share a common clock, and thus additional signaling is needed for
	      the decoder to be able to assign the received bits to the decoded symbols
	      unambiguously.
\end{itemize}
The main difference between the two settings is that, in the synchronous setting, the
encoder can use the absence of transmission---i.e., the transmission of a message $M_t$
with zero bits---as a signaling mechanism, while in the asynchronous case this is not
possible.

\section{Online Conformal Compression}
In this section, we introduce \gls{occ}, a zero-delay lossy compression scheme that leverages pre-trained sequence models to efficiently compress sequences while satisfying  the distortion requirement~\eqref{eqn:misc_guarantee1} for every user-defined distortion tolerance level $\alpha\in[0,1]$.
We start by reviewing \gls{ocp}~\cite{gibbs2021adaptive}.


\subsection{Online Conformal Prediction}
\Gls{ocp} is a calibration method for sequential decision-making processes that transforms the sequence of predictions of every black-box predictor into a sequence of prediction sets with long-term deterministic coverage guarantees.
Using the output $\mu_t(X|\hat{X}^{t-1})$  of an autoregressive predictor, at each time $t$, \gls{ocp} produces the prediction set
\begin{align}
	\label{eqn:conformal_set}
	\hat{\XXX}_{t}=\{X\in\XXX:\mu_t(X|\hat{X}^{t-1})\geq  \hat{Q}_t(1-\gamma_{t})\},
\end{align}
which includes all values $X\in\XXX$ whose predictive probability $\mu_t(X|\hat{X}^{t-1})$
crosses a threshold  $\hat{Q}_t(1-\gamma_{t})$. The reason for using the reconstruction
sequence $\hat{X}^{t-1}$ in \eqref{eqn:conformal_set}, instead of the actual sequence
$X^{t-1}$ will be made clear in the next subsection.

The threshold $\hat{Q}_t(1-\gamma_{t})$ corresponds to the $(1-\gamma_t)$ empirical quantile of the  predictive  probabilities assigned to the past symbols $\hat{X}^{t-1}$, i.e.,
\begin{multline} \label{eqn:quantile_definition}
	\hat{Q}_t(1-\gamma_t) =\\ \sup \left\{ p : \frac{1}{t-1} \sum_{s=1}^{t-1} \one{\{\mu_t(X_s|\hat{X}^{s-1})\geq p \}} \geq 1-\gamma_t  \right\}.
\end{multline}
Accordingly, the quantile level $1-\gamma_t$ determines the size of the prediction set \eqref{eqn:conformal_set}, with a larger $1-\gamma_t$ yielding a larger set.

For a given target miscoverage level   $\alpha$, \gls{ocp} updates the quantile level
online, based on error feedback about past predictions as
\begin{equation} \label{eqn:gamma_update}
	\gamma_{t+1}=\gamma_{t}-\eta_t(\one\{{X_{t}\notin \hat{\XXX}_{t}}\}-\alpha),
\end{equation}
where $\eta_t>0$ is a step size.
Therefore, when $X_{t}$ is not in the set $\hat{\mathcal{X}}_{t}$, the quantile level $1-\gamma_{t}$ is increased by $\eta_t(1-\alpha)$, causing future prediction sets to have larger sizes.
In contrast, if the prediction set $\hat{\mathcal{X}}_{t}$ includes $X_t$, the quantile level $1-\gamma_{t}$  is reduced by $\eta_t \alpha$, yielding smaller future prediction sets.

Following this adaptation strategy, \gls{ocp} is guaranteed to output prediction sets that satisfy the following long-term coverage guarantee.

\begin{lemma}
	\label{lemma:coverage}
	For every sequence $X_1,X_2,\dots$, every fixed value $\gamma_1\in[0,1]$ and every learning rate  $\eta_t=\eta_1 t^{-\beta}$ with $\beta\in[0,1)$, the \gls{ocp} prediction sets $\hat{\XXX}_{1},\hat{\XXX}_{2},\dots,$ satisfy the coverage guarantee
	\begin{align}
		\label{eqn:misc_guarantee}
		\left|\frac{1}{T}\sum^T_{t=1}\one{\{X_{t}\notin \hat{\XXX}_{t}\}}-\alpha\right|\leq \frac{1+\eta_1}{\eta_1 T^{1-\beta}}.
	\end{align}
\end{lemma}
\begin{proof}
	See~\cite{angelopoulos2024online}.
\end{proof}

Lemma \ref{lemma:coverage} shows  that, using \gls{ocp},  for every sequence of symbols
$X_1,X_2,\dots$ and target miscoverage $\alpha$, one can convert the output of
the prediction model $\mu_t(X_t|\hat{X}^{t-1})$ into prediction sets $\hat{\XXX}_{t}$ with
a miscoverage error ratio that converges to the nominal rate $\alpha$.

In the following, we show how to leverage the coverage properties of \gls{ocp} to devise a compression algorithm with a controllable distortion level.

\subsection{Synchronous Transmission}
By the assumption of noiseless transmission, at each time~$t$, both encoder and decoder have access to the reconstructed sequence $\hat{X}^{t-1}$, and hence to the predictive distribution $\mu_t(X_t|\hat{X}^{t-1})$.
Using this predictive distribution, both encoder and decoder evaluate the \gls{ocp} prediction set {$\hat{\XXX}_{t}$} in
(\ref{eqn:conformal_set}).
The key principle underlying the operation of \gls{occ} is to  allow for outage events only when the prediction set {$\hat{\XXX}_{t}$} does not include the current symbol $X_t$, so as to benefit from the guarantee (\ref{eqn:misc_guarantee}) in order to ensure the requirement (\ref{eqn:misc_guarantee1}).

To elaborate, consider first the synchronous transmission scenario. In this case,  if the current source symbol $X_t$ is in the prediction set $\hat{\XXX}_{t}$, the encoder applies a lossless entropy code to the \emph{truncated} distribution
\begin{align}
	\label{eqn:true_symbol_normalized_distribution}
	\bar{\mu}_t(x) & = \frac{\mu_t(x|\hat{X}^{t-1})}{\sum_{x'\in\hat{\XXX}_{t}}\mu_t(x'|\hat{X}^{t-1})} , ~~ \forall x \in \hat{\XXX}_t,
\end{align}
which is obtained by limiting the predictive distribution to the prediction set $\hat{\XXX}_t$.
This allows for the lossless reconstruction of $X_{t}$, i.e., $\hat{X}_t=X_t$.
Note that, thanks to the available common clock, there is no need for the code to be prefix-free~\cite{kontoyiannis2013optimal}.
In contrast, when $X_t\notin\hat{\XXX}_{t}$, the encoder signals the outage event by not transmitting.
In this case, the encoder and decoder agree to set the reconstructed symbol as
\begin{align}
	\label{eqn:symbol_out}
	\hat{X}_t=\arg\max_{x\notin\hat{\XXX}_{t}} \mu_t(x|\hat{X}^{t-1}).
\end{align}
Note that this is required to maintain the predictors at the encoder and at the decoder
synchronized.

\subsection{Asynchronous Transmission}
Compared to the synchronous scheme, where outages are implicitly signaled by the absence of a transmission, the asynchronous scheme explicitly encodes outage events.
To this end, we consider the augmented alphabet $\tilde{\XXX}_t =  \hat{\XXX}_t \cup \{\EEE\}$, where $\EEE$ is a symbol used to declare the outage event when $X_{t} \notin \hat{\mathcal{X}}_{t}$.
If an outage occurs, the reconstructed symbol is selected as specified in \eqref{eqn:symbol_out}.
In contrast to the synchronous case, in order to enable parsing at the decoder, 
the encoder employs a lossless prefix-free entropy
code to encode~$X_t$.
Specifically, $X_{t}$ is treated as a sample from the augmented truncated
distribution
\begin{align}
	\label{eqn:true_symbol_normalized_distribution_asynch}
	\tilde{\mu}_t(x)    & = \bar{\mu}_t(x) (1-\alpha) , ~~ \forall x \in \hat{\XXX}_t, \\
	\label{eqn:erasure_distribution}
	\tilde{\mu}_t(\EEE) & = \alpha.
\end{align}

In the Appendix, we show that assigning a probability
$\tilde{\mu}_t(\EEE) = \alpha$ to the outage symbol is the asymptotically optimal
\emph{non-adaptive} strategy to minimize the transmission
rate~\eqref{eqn:avg_compression_rate}.

\subsection{Distortion Guarantee}
We now derive an upper bound on the distortion of the \gls{occ} scheme, showing that it
satisfies the distortion requirement~\eqref{eqn:misc_guarantee1}.
By construction, whenever $X_t\in\hat{\XXX}_t$,
the reconstructed symbol $\hat{X}_t$ coincides with the source symbol $X_t$. Therefore,
the distortion error on every sequence $X^{T}$ can be readily upper-bounded by the average
number of miscoverage errors as
\begin{equation}
	\frac{1}{T}\sum_{t=1}^{T} \one{\{\hat{X}_t \neq X_t\}}\leq\frac{1}{T}\sum_{t=1}^{T} \one{\{X_t \notin \hat{\XXX}_t\}}.
\end{equation}
By Lemma \ref{lemma:coverage}, we have the following result.
\begin{theorem}\label{theorem:outage_guarentee}
	For every learning rate sequence $\eta_t=\eta_1 t^{-\beta}$ with $\beta\in[0,1)$,
	the average distortion of \gls{occ} on every sequence $X^{T}$ can be bounded as
	\begin{equation}\label{eqn:outage_upper_bound}
		\frac{1}{T} \sum_{t=1}^{T} \one{\{\hat{X}_t \neq X_t\}} \leq \alpha + \frac{1+\eta_1}{\eta_1 T^{1-\beta}}.
	\end{equation}
\end{theorem}

\section{Results}
In this section, we evaluate the performance of \gls{occ} for the task of compressing English text, and we benchmark the proposed schemes against state-of-the-art prediction-based compression schemes.

\subsection{Setting}
We consider compressing English text from Shakespeare’s works and Taylor Swift’s songs using datasets from \cite{karpathy2015charrnn} and \cite{huggingartists}, respectively, along with a sequence predictor implemented via a \gls{llm}.
First,  the text is tokenized, 
and the distribution of the next token is modeled using the nanoGPT-$2$ model,\footnote{https://github.com/karpathy/nanoGPT} which is fine-tuned on a holdout portion of the Shakespeare dataset of size $\num{301966}$ tokens.
The nanoGPT-$2$ model operates on a vocabulary comprising $|\mathcal{X}|=\num{50257}$ different tokens, which correspond to the alphabet of the symbols to be encoded.
As such, the  uncompressed transmission of the sequence of tokens requires $16$ bits per token.
For the simulations, we have used an initial threshold $\gamma_1 = \alpha$ and a constant learning parameter $\eta_t=0.001$.

\subsection{Benchmarks}
In our experiments, we evaluate the following benchmark compression schemes.

\subsubsection{Dropout-LLMZip} LLMZip~\cite{valmeekam2023llmzip} is a lossless compression scheme that encodes each token $X_t$  by using an entropy coding scheme based on the model's predictive distribution $\mu_t(X_t|{X}^{t-1})$.
To ensure that the constraint~\eqref{eqn:dist_guarantee} is satisfied, we consider a variant of LLMZip, in which, at each round $t$, the encoder declares an outage independently with probability~$\alpha$. When an outage occurs, the transmitter and receiver agree to set $\hat{X}_t=\argmax_{x\in\mathcal{X}}\mu_t(x|\hat{X}^{t-1})$.
This approach is referred to as Dropout-LLMZip.



\subsubsection{Block Conformal Compression}
We consider a block-based variant of \gls{occ}, where the encoder has access to the entire sequence $X^T$ before encoding.
This scheme, referred to as \gls{bcc}, selects the largest $\gamma_t = \gamma^*$ in the prediction set~\eqref{eqn:conformal_set} that satisfies the distortion requirement~\eqref{eqn:misc_guarantee1}.
\Gls{bcc} serves as a benchmark to evaluate the ability of \gls{occ} to adapt online, mimicking an ideal static threshold selection strategy, with threshold chosen in hindsight.





\subsection{Results}


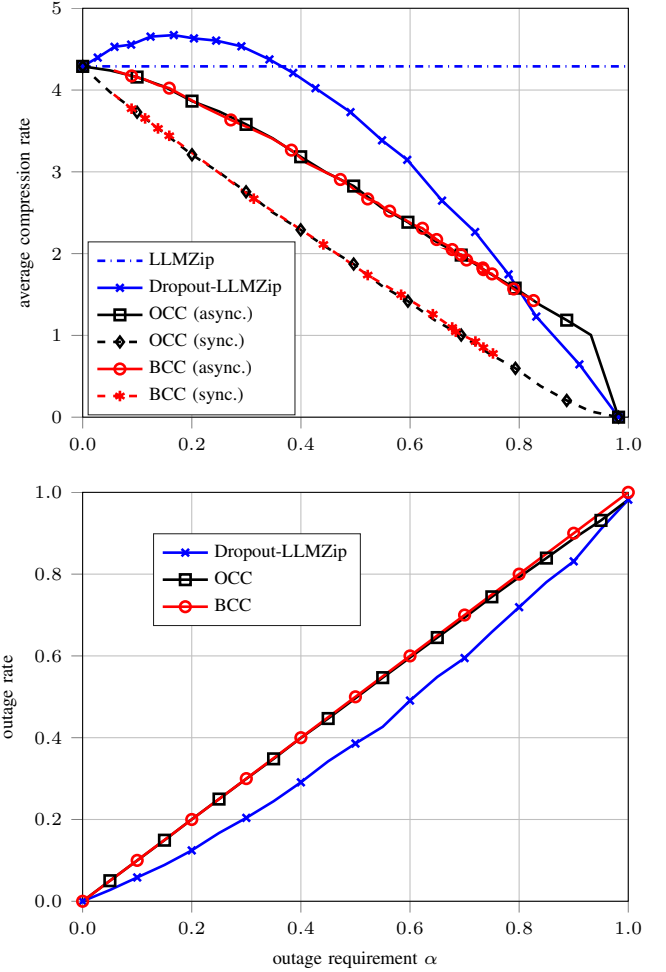
\begin{figure}[!t]
	\centering
%
%
\begin{tikzpicture}[font=\scriptsize]

\begin{groupplot}[group style={group size=1 by 2}]
\nextgroupplot[
width=0.4\textwidth,
height=0.3\textwidth,
scale only axis,
tick align=outside,
tick pos=left,
xmajorgrids,
xmin=0, xmax=1,
xtick={0,0.2,0.4,0.6,0.8,1},
xticklabels={
  \(\displaystyle {0.0}\),
  \(\displaystyle {0.2}\),
  \(\displaystyle {0.4}\),
  \(\displaystyle {0.6}\),
  \(\displaystyle {0.8}\),
  \(\displaystyle {1.0}\)
},
ylabel={average compression rate},
ymajorgrids,
ymin=0, ymax=5,
ytick={0,1,2,3,4,5},
yticklabels={
  \(\displaystyle {0}\),
  \(\displaystyle {1}\),
  \(\displaystyle {2}\),
  \(\displaystyle {3}\),
  \(\displaystyle {4}\),
  \(\displaystyle {5}\)
},
legend style={legend cell align=left, align=left, draw=white!15!black, legend style={at={(0.39,0.43)}}}
]
\addplot [dashdotted, color=blue, line width = 1.0]
  table{%
0 4.28954285714286
1 4.28954285714286
};
\addlegendentry{LLMZip}

\addplot [color=blue, line width=1.0pt, mark=x, mark options={solid}]
table {%
0 4.28954285714286
0.0275428571428571 4.397
0.0582571428571429 4.52842857142857
0.0888 4.55551428571429
0.124371428571429 4.65311428571428
0.166971428571429 4.67165714285714
0.203942857142857 4.63151428571429
0.244628571428571 4.60448571428571
0.290885714285714 4.53477142857143
0.342285714285714 4.37434285714286
0.385771428571429 4.20862857142857
0.426657142857143 4.02257142857143
0.490828571428572 3.72971428571429
0.548628571428572 3.38517142857143
0.594771428571429 3.14662857142857
0.6586 2.64754285714286
0.719171428571429 2.2642
0.780514285714286 1.74708571428571
0.831285714285715 1.2302
0.910285714285715 0.645742857142857
0.982085714285715 0
};
\addlegendentry{Dropout-LLMZip}

\addplot [color=black, line width=1.0pt, mark=square, mark options={solid}, mark repeat=2, mark phase=1]
table {%
0 4.28954285714286
0.0502857142857143 4.23968571428571
0.0996571428571429 4.15877142857143
0.149342857142857 4.0424
0.200314285714286 3.86465714285714
0.249914285714286 3.7384
0.299428571428571 3.58054285714286
0.348285714285714 3.40977142857143
0.399514285714286 3.18417142857143
0.446914285714286 2.98942857142857
0.4966 2.8262
0.546857142857143 2.56745714285714
0.596028571428571 2.3838
0.644914285714286 2.15725714285714
0.693685714285714 1.98225714285714
0.744485714285714 1.76371428571429
0.792542857142857 1.57817142857143
0.839228571428572 1.36854285714286
0.887057142857143 1.18628571428571
0.931371428571429 1.00517142857143
0.982085714285715 0
};
\addlegendentry{\gls{occ} (async.)}

\addplot [dashed, color=black, line width=1.0pt, mark=diamond, mark options={solid}, mark repeat=2, mark phase=1]
table {%
0 4.28954285714286
0.0502857142857143 3.97442857142857
0.0996571428571429 3.72914285714286
0.149342857142857 3.47351428571429
0.200314285714286 3.20888571428571
0.249914285714286 2.99417142857143
0.299428571428571 2.75585714285714
0.348285714285714 2.50328571428571
0.399514285714286 2.291
0.446914285714286 2.07625714285714
0.4966 1.87251428571429
0.546857142857143 1.61242857142857
0.596028571428571 1.41882857142857
0.644914285714286 1.18694285714286
0.693685714285714 1.00405714285714
0.744485714285714 0.779285714285714
0.793057142857143 0.5976
0.839228571428572 0.380085714285714
0.887057142857143 0.2048
0.931371428571429 0.0706285714285715
0.982085714285715 0
};
\addlegendentry{\gls{occ} (sync.)}

\addplot [color=red, line width=1.0pt, mark=o,mark options={solid}, mark repeat=3, mark phase=2]
table {%
0.057 4.23905714285714
0.0896 4.17257142857143
0.114428571428571 4.12834285714286
0.138 4.06048571428571
0.158685714285714 4.02331428571429
0.158685714285714 4.02331428571429
0.221657142857143 3.79442857142857
0.271485714285714 3.63442857142857
0.313542857142857 3.51145714285714
0.346457142857143 3.40977142857143
0.382542857142857 3.2654
0.409685714285714 3.11114285714286
0.441114285714286 3.00717142857143
0.472285714285714 2.9064
0.491514285714286 2.81854285714286
0.503628571428572 2.74917142857143
0.522342857142857 2.66845714285714
0.541142857142857 2.61871428571429
0.551828571428572 2.56294285714286
0.562542857142857 2.51985714285714
0.583571428571429 2.44448571428571
0.606228571428572 2.3518
0.622685714285715 2.31014285714286
0.634228571428572 2.24948571428571
0.641714285714286 2.21094285714286
0.648657142857143 2.17222857142857
0.647657142857143 2.17057142857143
0.667657142857143 2.10242857142857
0.6772 2.04971428571429
0.674857142857143 2.0428
0.671285714285714 2.04542857142857
0.693485714285714 1.98648571428571
0.683857142857143 1.99417142857143
0.711942857142857 1.91151428571429
0.703542857142857 1.92165714285714
0.717085714285715 1.88714285714286
0.719714285714286 1.883
0.733342857142857 1.82482857142857
0.728942857142857 1.83842857142857
0.7294 1.81942857142857
0.7344 1.80442857142857
0.738885714285715 1.78568571428571
0.736514285714286 1.78242857142857
0.7502 1.75194285714286
0.751914285714286 1.73808571428571
0.769 1.65494285714286
0.789914285714286 1.5676
0.796942857142857 1.5336
0.8212 1.45137142857143
0.826257142857143 1.42591428571429
};
\addlegendentry{\gls{bcc} (async.)}

\addplot [dashed, red, line width=1.0pt, mark=asterisk,mark options={solid}, mark repeat=1, mark phase=2]
table {%
0.057 3.94622857142857
0.0896 3.7718
0.114428571428571 3.65322857142857
0.138 3.53094285714286
0.158685714285714 3.44071428571429
0.313542857142857 2.67154285714286
0.441114285714286 2.11191428571429
0.522342857142857 1.73777142857143
0.583571428571429 1.4972
0.641714285714286 1.25911428571429
0.6772 1.09682857142857
0.683857142857143 1.0444
0.719714285714286 0.9252
0.7344 0.851085714285714
0.751914285714286 0.779142857142857
};
\addlegendentry{\gls{bcc} (sync.)}

\nextgroupplot[
width=0.4\textwidth,
height=0.3\textwidth,
scale only axis,
tick align=outside,
tick pos=left,
xlabel={outage requirement $\alpha$},
xmajorgrids,
xmin=0, xmax=1,
xtick={0,0.2,0.4,0.6,0.8,1},
xticklabels={
  \(\displaystyle {0.0}\),
  \(\displaystyle {0.2}\),
  \(\displaystyle {0.4}\),
  \(\displaystyle {0.6}\),
  \(\displaystyle {0.8}\),
  \(\displaystyle {1.0}\)
},
ylabel={outage rate},
ymajorgrids,
ymin=0, ymax=1,
ytick={0,0.2,0.4,0.6,0.8,1},
yticklabels={
  \(\displaystyle {0.0}\),
  \(\displaystyle {0.2}\),
  \(\displaystyle {0.4}\),
  \(\displaystyle {0.6}\),
  \(\displaystyle {0.8}\),
  \(\displaystyle {1.0}\)
},
legend style={legend cell align=left, align=left, draw=white!15!black, legend style={at={(0.51,0.9)}}}
]
\addplot [color=blue, line width=1.0pt, mark=x, mark options={solid}, mark repeat=2]
table {%
0 0
0.05 0.0275428571428571
0.1 0.0582571428571429
0.15 0.0888
0.2 0.124371428571429
0.25 0.166971428571429
0.3 0.203942857142857
0.35 0.244628571428571
0.4 0.290885714285714
0.45 0.342285714285714
0.5 0.385771428571429
0.55 0.426657142857143
0.6 0.490828571428572
0.65 0.548628571428572
0.7 0.594771428571429
0.75 0.6586
0.8 0.719171428571429
0.85 0.780514285714286
0.9 0.831285714285715
0.95 0.910285714285715
1 0.982085714285715
};
\addlegendentry{Dropout-LLMZip}

\addplot [color=black, line width=1.0pt, mark=square, mark options={solid}, mark repeat=2, mark phase=2]
table {%
0 0
0.05 0.0502857142857143
0.1 0.0996571428571429
0.15 0.149342857142857
0.2 0.200314285714286
0.25 0.249914285714286
0.3 0.299428571428571
0.35 0.348285714285714
0.4 0.399514285714286
0.45 0.446914285714286
0.5 0.4966
0.55 0.546857142857143
0.6 0.596028571428571
0.65 0.644914285714286
0.7 0.693685714285714
0.75 0.744485714285714
0.8 0.792542857142857
0.85 0.839228571428572
0.9 0.887057142857143
0.95 0.931371428571429
1 0.982085714285715
};
\addlegendentry{\gls{occ}}


\addplot [color=red, line width=1.0pt, mark=o, mark options={solid}, mark repeat=2]
table {%
0 0
0.05 0.05
0.1 0.1
0.15 0.15
0.2 0.2
0.25 0.25
0.3 0.3
0.35 0.35
0.4 0.4
0.45 0.45
0.5 0.5
0.55 0.55
0.6 0.6
0.65 0.65
0.7 0.7
0.75 0.75
0.8 0.8
0.85 0.85
0.9 0.9
0.95 0.95
1 1
};
\addlegendentry{\gls{bcc}}

\end{groupplot}

\end{tikzpicture}%
	\caption{Average compression rate and outage rate versus the outage requirement $\alpha$.}
	\label{fig:average_compression_rate_vs_outage}
\end{figure}

In Fig.~\ref{fig:average_compression_rate_vs_outage}, we report the average compression rate~\eqref{eqn:avg_compression_rate} and the distortion~\eqref{eqn:misc_guarantee1} for different outage requirements $\alpha$.
The results are averaged over $10$ different sequences of length $T = 3500$ from the Shakespeare’s corpus.
We observe that \gls{occ} delivers substantial compression rate gains as $\alpha$ increases.
For example, compared to Dropout-LLMZip, for $\alpha = 0.2$, \gls{occ} reduces the compression rate by $30\%$ in the synchronous case and by $16\%$ in the asynchronous case,  under the same distortion requirement.
Note that the compression rate of Dropout-LLMZip is worse than that of the distortion-free
case $\alpha=0$ for values of $\alpha$ up to $0.35$. This performance degradation is due to
the introduction of errors in the reconstructed sequence, which affect the quality of the
model’s predictive distribution, and lead to a larger compression rate.

All the schemes perform better under the synchronous scenario due to the absence of outage signaling overhead.
Furthermore, when comparing the performance of the \gls{occ} against \gls{bcc}, we observe that the two deliver essentially the same compression rate.
This indicates that \gls{occ} is competitive with a block-based method that is allowed to choose the predictive set threshold in hindsight.
Finally, all schemes are seen to comply with the distortion requirement, as they deliver an outage rate lower than the target level~$\alpha$.
However, it is important to note that the outage guarantees of  Dropout-LLMZip only hold on average, but not on a per-sequence basis.

\begin{figure}[!t]
	\centering
	\input{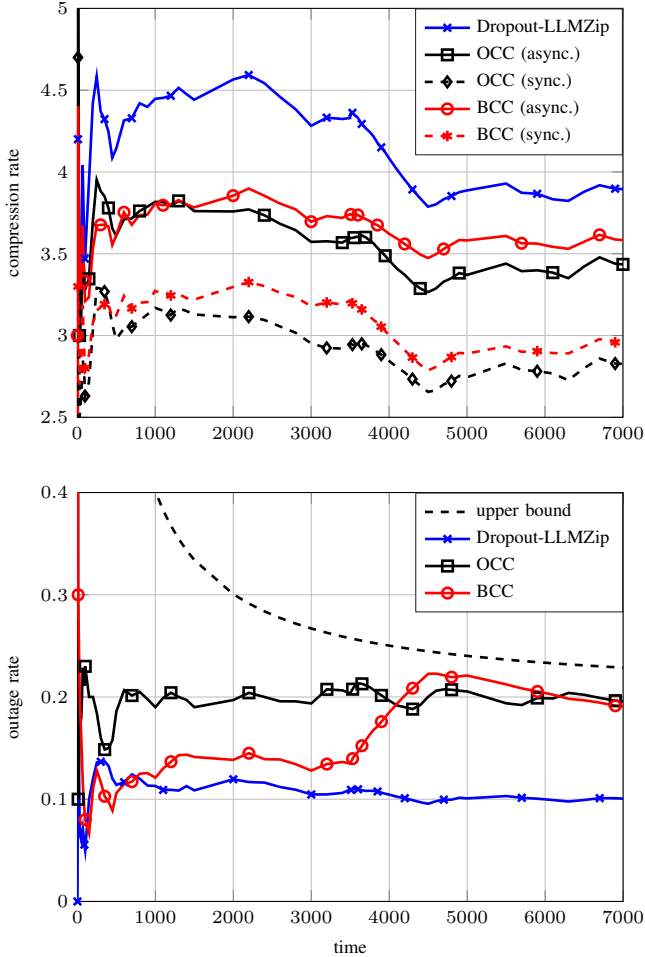}
	\caption{Average compression rate and outage rate versus the time index $t$.}
	\label{fig:compression_outage_time}
\end{figure}

We now consider a non-stationary scenario in which the data distribution changes over time.
Specifically, we consider a sequence of length $T=7000$ in which the first half of the symbols correspond to tokens from the Shakespeare validation dataset, while the remaining half of the tokens are from a Taylor Swift song.
Recall that the predictive model is fine-tuned on the Shakespeare dataset only.

In Fig.~\ref{fig:compression_outage_time}, we report the compression rate and outage rate as a function of time.
We observe that \gls{occ} delivers a compression rate that is substantially lower than Dropout-LLMZip and comparable to the rate obtained by \gls{bcc}.
Note that, although adaptive, \gls{occ} delivers an outage rate that remains uniform over time.
In contrast, \gls{bcc}, being allowed to see the entire sequence before encoding, yields a smaller distortion over the Shakespeare data as compared to the Taylor Swift dataset, for which the model is not fine-tuned.





\glsresetall
\section{Conclusion}
In this paper, we introduced \gls{occ}, a novel zero-delay lossy source compression scheme that leverages pre-trained sequence predictors at the encoder and decoder.
\Gls{occ} guarantees deterministic target outage levels at any time.
Furthermore, numerical results have demonstrated that \gls{occ} achieves a compression
rate comparable to that of an offline scheme, in which the encoder, with hindsight,
selects the optimal subset of symbols to satisfy the outage constraints. These findings
highlight the effectiveness of \gls{occ} in achieving adaptive, efficient, and reliable
lossy compression in an online setting.


\appendix[Asymptotically Optimal Non-Adaptive Strategy]
\label{app:choice_erasure}
We show that assigning a probability  $\tilde{\mu}_t(\EEE)=\alpha$ to the outage symbol $\EEE$ in the definition of the truncated distribution $\tilde{\mu}_t$ defined in~\eqref{eqn:true_symbol_normalized_distribution_asynch}--\eqref{eqn:erasure_distribution} is the asymptotically optimal non-adaptive strategy.

Encoding a sequence of symbols $X^T$ using an optimal prefix-free entropy code, such as the Huffman code, and assuming at each time $t$ the symbol $X_{t}$ is treated as sample from the augmented truncated distribution $\tilde{\mu}_t(x)=\bar{\mu}_{t}(x)(1-\epsilon)$ for $x\in \hat{\mathcal{X}}_t$, and $\tilde{\mu}_{t}(\EEE)=\epsilon$, we obtain the
following  upper bound on the  transmission rate~\cite{thomas2006elements}:
\begin{IEEEeqnarray}{rCl}
	\frac{1}{T}  \sum^T_{t=1}b_t &\leq& 1-\frac{1}{T}\sum^T_{t=1}\mathbbm{1}{\{X_t \in \hat{\mathcal{X}}_t\}}\log_2(\bar{\mu}_t(X_t)) \nonumber\\
	&&- \frac{1}{T}\sum^T_{t=1}\biggl[\mathbbm{1}{\{X_t \notin \hat{\mathcal{X}}_t\}}\log_2(\epsilon) \nonumber \\
		&&+ \mathbbm{1}{\{X_t \in \hat{\mathcal{X}}_t\}}\log_2(1-\epsilon)\biggr].\label{eq:uppebound}
\end{IEEEeqnarray}
%
By virtue of Lemma~\ref{lemma:coverage}, the average number of miscoverage events converges to~$\alpha$,
\begin{align}
	\lim_{T\to\infty} \frac{1}{T}\sum^T_{t=1}\mathbbm{1}{\{X_t \notin \hat{\mathcal{X}}_t\}} = \alpha,
\end{align}
and therefore, for $T\to\infty$, the value of $\epsilon$ that minimizes the upper bound \eqref{eq:uppebound} is $\epsilon = \alpha$.

\bibliographystyle{IEEEtran}
\bibliography{references}

\end{document}